\documentclass{KMVC}
\hypersetup{
    colorlinks,
    linkcolor={red!50!black},
    citecolor={blue!50!black},
    urlcolor={blue!80!black}
}

\usepackage{subcaption}
\usepackage[bitstream-charter]{mathdesign}
\DeclareSymbolFont{usualmathcal}{OMS}{cmsy}{m}{n}
\DeclareSymbolFontAlphabet{\mathcal}{usualmathcal}

\fancypagestyle{SPstyle}{
\fancyhf{}

\fancyfoot[C]{\textbf{\thepage}}
}
\usepackage{enumitem}
\usepackage{multicol}
\newcommand{\parallelsum}{\mathbin{\!/\mkern-5mu/\!}}
\newenvironment{psmallmatrix}
  {\left(\begin{smallmatrix}}
  {\end{smallmatrix}\right)}

\begin{document}

\pagestyle{SPstyle}

\begin{center}{\Large \textbf{\color{darkdeepblue}{
The effect of Coulomb interactions on relic neutrino detection via beta decaying impurities in (semi)metals\\
}}}\end{center}

\begin{center}\textbf{
Karel van der Marck\textsuperscript{1,2$\star$} and
Vadim Cheianov\textsuperscript{1$\dagger$}
}\end{center}

\begin{center}
{\bf 1} Instituut-Lorentz, Universiteit Leiden, Niels Bohrweg 2, 2333 CA Leiden, The Netherlands
\\
{\bf 2} Institute for Theoretical Physics, Universiteit van Amsterdam, Science Park 904, 1090 GL, The Netherlands
\\[\baselineskip]
$\star$ \href{mailto:email1}{\small karel.vandermarck@ziggo.nl}\,,\quad
$\dagger$ \href{mailto:email2}{\small cheianov@lorentz.leidenuniv.nl}
\end{center}

\section*{\color{darkdeepblue}{Abstract}}
\textbf{\boldmath{
\hspace{-0.15cm}Measuring the electron neutrino mass is a long-standing objective and requires a high energy resolution of certain $\beta$-decay experiments, as well as a visible cosmic neutrino background ($C\hspace{-0.05cm}\nu B$) spectrum. Many quantum mechanical and chemical effects could potentially impair the required resolution/visibility, e.g., the Coulomb interactions between the electrons in the $\beta$-decaying impurity and in the solid-state environment. We analyze the effect when hybridization is suppressed completely using a dielectric spacer, and also when hybridization is present up to the lowest nontrivial order in perturbation theory.
}}

\vspace{\baselineskip}

\vspace{10pt}
\noindent\rule{\textwidth}{1pt}
\tableofcontents
\noindent\rule{\textwidth}{1pt}
\vspace{10pt}

\section{Introduction}
\label{sec:intro}
\vspace{-0.3cm}

The discovery of neutrino oscillations is a milestone in particle physics, as it conclusively demonstrates that neutrinos are massive particles. This result follows from several experiments by the Super-Kamiokande collaboration \cite{SuperKamiokande} together with the findings from the group at the Sudbury Neutrino Observatory \cite{SNO}. The leaders of these groups, T. Kajita and A. McDonald, respectively, were awarded the Nobel Prize in Physics in 2015 for their contributions to the aforementioned experiments.

Oscillation experiments gave access to the relative values of the three mass eigenstates of neutrinos. An ambitious next goal would be to measure the absolute masses. A first proposal to realize this objective dates back to a 1962 paper in which S. Weinberg suggested that the mass of an electron neutrino can be determined by investigating the incoming and outgoing energies of $\beta$-decay \cite{Weinberg}. Either an electron antineutrino is an outgoing particle, or an electron neutrino is an incoming particle:

\vspace{-0.3cm}
\begin{subequations}
\begin{align}
&^{A}_{Z}X \hspace{1.07cm}\to\hspace{0.3cm} ^{\hspace{0.33cm}A}_{Z+1}X + e + \overline{\nu}_{e},\label{First reaction}\\
&^{A}_{Z}X + \nu_{e} \hspace{0.3cm}\to\hspace{0.3cm} ^{\hspace{0.33cm}A}_{Z+1}X + e.\label{Second reaction}
\end{align}
\label{Two reactions}
\end{subequations}
\vspace{-0.3cm}

\begin{figure}[h]
    \centering
    \includegraphics[width=0.57\textwidth]{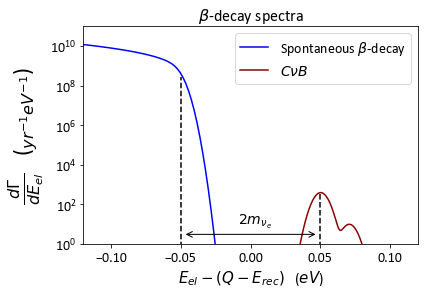}
    \caption{The spontaneous $\beta$-decay has a continuous spectrum, for the electron's energy is measured and the energy in channel \ref{First reaction} can be distributed among the kinetic energy of the electron and the kinetic energy of the electron antineutrino. In channel \ref{Second reaction}, however, the $C\hspace{-0.05cm}\nu B$ is captured and all energy is transformed into the motion of the electron, making the spectrum peaked around the neutrino masses. To make this plot, we have assumed that $m_{\nu_{e}} = 0.05$ $eV$.}
    \label{fig: beta decay spectra}
\end{figure}

The equations above lead to two disconnected regions of the $\beta$-spectrum separated by $2m_{\nu_{e}}c^{2}$, exactly twice the mass of the electron neutrino, see Fig. \ref{fig: beta decay spectra} \cite{FundamentalProblemsRelicNeutrinoDetection}. Reaction \ref{First reaction} can take place spontaneously, whereas we need a neutrino source to realize stimulated emission, channel \ref{Second reaction}. In Weinberg's proposal, this would be the cosmic neutrino background ($C\hspace{-0.05cm}\nu B$). Unlike the cosmic microwave background ($CMB$), which separated from matter $\sim380\hspace{0.05cm}000$ years after the Big Bang, the $C\hspace{-0.05cm}\nu B$ decoupled much earlier: it was produced within the first second after the Big Bang. On this account, the $C\hspace{-0.05cm}\nu B$ can help us better understand the recombination age, and can therefore be studied for various purposes \cite{Nicola}.

The upper bound of the electron neutrino mass limits the possibilities for experimental implementations of Weinberg's idea quite substantially, since the lower the electron neutrino mass, the higher the required energy resolution \cite{KATRIN,ExtendingLCDM}.

The neutrino mass could potentially be determined using only the spontaneous $\beta$-decay channel. This removes the difficult objective of measuring relic neutrinos and has proven to be successful when the KATRIN collaboration found an upper limit of $0.85$ $eV$ in 2022\cite{KATRIN} and of $0.45$ $eV$ in 2025\cite{KATRIN2} using ultra-cold molecular tritium gas. Nevertheless, we will still consider the $C\hspace{-0.05cm}\nu B$ channel in this work, as its future measurement could improve the precision of the neutrino mass determination. The reason lies in the limited event rate near the endpoint of the spontaneous beta spectrum, which is the region most sensitive to the neutrino mass. A detection of the $C\hspace{-0.05cm}\nu B$ would provide an independent and complementary probe.


A particularly relevant realization of Weinberg's idea is provided by tritium atoms adsorbed on graphene, which constitutes the target material proposed in the PTOLEMY experiment.\cite{PTOLEMY} In this configuration the $\beta$-decay occurs for tritium bound to a two-dimensional host lattice rather than in molecular form. The electronic environment of the decay therefore differs qualitatively from the molecular tritium case traditionally considered in neutrino-mass experiments. The graphene substrate meets several key requirements: atomic form tritium, spatial precision, target volume minimization, environmental safety, and minimal energy loss \cite{PTOLEMY2}. Besides, the graphene sheet is a conductor when it is not fully hydrogenated, which is essential to prevent broadening of the $\beta$-spectrum due to an inhomogeneous charge distribution.

The PTOLEMY concept relies on tritium adsorbed on graphene, since it has a large neutrino capture cross-section and the emitted electron has a low decay energy. Yet, there are several solid-state effects that could impair the visibility of the relic neutrino spectrum due to the quantum induced broadening of the $\beta$-spectrum \cite{FundamentalProblemsRelicNeutrinoDetection,FundamentalProblemsRelicNeutrinoDetection2,PTOLEMY3,Nussinov,TanCheianov,GrapheneTritiumInteractionDFT}. One of which is the zero-point motion of the $\beta$-decayer, which rules out light isotopes such as tritium. Some suggested alternatives are $^{63}Ni$, $^{151}Sm$, $^{171}Tm$, and $^{241}Pu$\cite{Plutonium_beta_decayer,FundamentalProblemsRelicNeutrinoDetection,FundamentalProblemsRelicNeutrinoDetection2}, although it should be noted that these options are not all selected based on practicability. One would need, for instance, a target mass of 350 kg $^{171}Tm$ for an equal neutrino capture cross section to that of 100 g of tritium\cite{EmpiricalResultsFeasibility} (and see \cite{NeutrinoCaptureCrossSectionCalculation} for the first calculation of said neutrino capture cross sections), while the high purity production remains a difficult problem.

Besides the zero-point motion, the Coulomb interactions between the impurity electron and the environment might affect the visibility of the $C\hspace{-0.05cm}\nu B$-spectrum. If the daughter ion has a finite lifetime, then the $\beta$-spectrum will obscure the $C\hspace{-0.05cm}\nu$B-signal due to the Lorentzian broadening \cite{TanCheianov}. One way to circumvent this effect, is to ensure that the daughter ion be stable against the capture of an electron from the solid-state environment. Such an approach requires some fine electro-chemical tuning as discussed in Section \ref{sec: E&M} of this paper. In this discussion, the stability is vastly dependent on the choice of isotope.

In an alternative scenario, both daughter and mother isotopes hybridize with the solid-state environment forming the so-called fractional charge states, that is states which are superpositions of pure charge states having different degrees of ionization. In this situation, the fractionality of the localized atomic charge ensures the possibility of a direct transition between the ground state of the mother isotope and the ground state of the daughter isotope through the $\beta$-decay process. Such a scenario is natural to all kinds of chemical bonding and does not require fine tuning. However, the visibility of the relic neutrino peak in the $\beta$-spectrum still requires the presence of an X-ray edge singularity in such a process. To the best of our knowledge, the X-ray edge has not been investigated in such a particular setting, therefore we present its analysis in Section \ref{sec: quantum}.

By accounting for the Coulomb interactions between the graphene and the impurity, we analyze a single solid-state effect on the resolution of the PTOLEMY project. The other effects are, inter alia, phonon emission and Friedel oscillations. The influence of these phenomena on relic neutrino detection requires further investigation.

\vspace{-0.3cm}
\section{Problem Setting}
\vspace{-0.3cm}

In Section \ref{sec: E&M}, we assume negligible coupling between the $\beta$-emitter and the graphene layer, so that the impurity states have integer-valued charge. This way, we can estimate the energy cost or gain from donating or receiving an electron from the solid-state environment based on the Coulomb potential in the impurity.

The Coulomb potential depends on the ionization energy of the impurity in a vacuum, but also on the graphene layer, which we will assume to be an infinite conducting layer. The method of image charges allows us to compute the contribution of the graphene layer to the potential, but the dielectric spacer is not yet incorporated. To complete the story, we insert an image of the slab, we remove the image charges and replace them with two infinite lines of equidistant image charges in the opposite half-space w.r.t. the impurity that reproduce the electric field experienced by the electron and satisfy the boundary conditions at the interface of the dielectric slab, Fig. \ref{fig: image charges} \cite{InfiniteImageCharges}. By summing up the contributions of all images, one can calculate the force on the electron and thus obtain the correction to the potential by integration.

Fortunately, the experiment enjoys some parameters that can be tuned to optimize for stability, such as the thickness of the dielectric layer, the distance between the layer and the $\beta$-emitter, and the work function of graphene. To see if this wiggle room is a remedy for a stable impurity and its daughter isotope, we set up a parameter space diagram, Fig. \ref{fig: stability plots} and we test the different candidates for suitable $\beta$-emitters, typically lanthanides such as thulium.

In Section \ref{sec: quantum}, we explore the stability prospects when the (semi)metal-impurity coupling is \textit{not} negligible. It follows that the charge states need not be integer-valued, resulting in multiple decay channels. These channels are encoded in the spectral function, which we compute in Section \ref{subsec: spectral function}. If this function is IR divergent with a one-sided power law, then this is called the \textit{X-ray edge singularity}, which can preserve the visibility of the $C\hspace{-0.05cm}\nu B$-spectrum \cite{Infinite_hole_mass,Many_electron_singularity,Nozieres_I,Nozieres_II,Nozieres_III,SchotteSchotte}.

The solid-state environment is modeled with massless Weyl fermionic fields, which we \textit{bosonize} in Section \ref{Bosonization}, resulting in a \textit{Tomonaga-Luttinger liquid} (TLL) \cite{TLL}. We insert an Anderson-type electronic impurity and treat the hybridization perturbatively. We study the presence of the X-ray edge singularity at the lowest nontrivial order term.

\vspace{-0.3cm}
\section{Classical E\&M Approach}
\label{sec: E&M}
\vspace{-0.3cm}

It is suggested that the visibility of the $C\hspace{-0.05cm}\nu B$ signal can be improved by maximizing the distance between the $\beta$-decayers and the graphene sheet with the insertion of a dielectric slab \cite{TanCheianov}. At first glance, several promising candidates present themselves, inter alia, metal-organic self-assembly layers \cite{Self_assembly_organic_metal_layers}, hexagonal boron nitride (h-BN) \cite{GrapheneDielectricH-BN,Metal_on_h-BN}, silicon dioxide (SiO$_2$) \cite{Metal_on_SiO2}, and polymethyl methacrylate (PMMA) \cite{Metal_on_polymer}, but rather than restricting our considerations to any particular geometry, we approach the problem classically with the method of image charges and estimate the size of quantum corrections at the end of Section \ref{sec: E&M}.

We intend to find out if we can have a stable $\beta$-emitter in the solid-state system and simultaneously obtain a stable daughter isotope. In other words, we want to look for configurations in which it is energetically unfavorable for both the $\beta$-decayer and the daughter isotope to donate or gain an electron from the graphene. We will assume integer-charged states, i.e., negligible coupling between the $\beta$-emitter and the graphene layer.

In a vacuum, one can simply compare the ionization energy with the work function $(W)$ for processes $A^{Q} \to A^{Q+1}$, and compare the electronic affinity with $W$ for processes $A^{Q} \to A^{Q-1}$. Note that $W$ may be tuned between $3.25$ and $5.54$ $eV$ \cite{GapheneWF}. For our purposes, the $\beta$-decayer will face a graphene layer separated by a dielectric substrate. This setup creates a potential felt by the electrons in the $\beta$-emitter, which we need to correct for. 

\subsection{Energy Potential}

\begin{figure*}[ht]
\centering
\centering \includegraphics[width=0.99\linewidth]{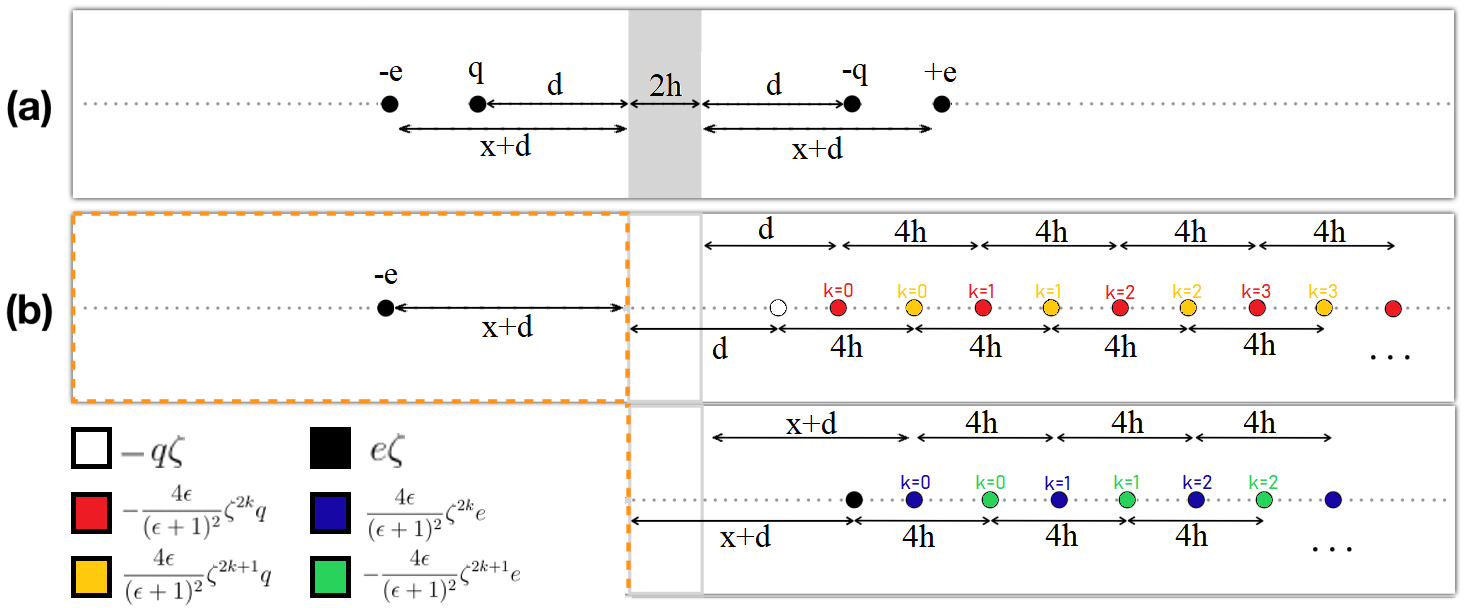}
\caption{\begin{tabular}[t]{rl}
a)& There are real charges $-e$ and $q$ on the left side of the graphene layer.\\
&The corresponding image charges are on the right side of the graphene\\
&layer. Also the substrate has a mirror image on the right side, which\\
&effectively makes it twice as thick.\\
b)& The charge $-e$ feels the electric field from the other three charges in Fig.\\
&\ref{fig: image charges}\color{red!50!black}a\color{black}. Equivalently, we remove these three charges and the substrate, and\\
&we replace them with an infinite number of image charges that generate\\
&the same field \cite{InfiniteImageCharges}. The charge values are found in the legend, bottom\\
&left of the figure.
\end{tabular}}
\label{fig: image charges}
\end{figure*}

Without the dielectric layer (thickness $h \to 0$), the image charge method tells us that the force felt by the electron is the sum of the forces due to (i) the nucleus, (ii) the image of the electron, and (iii) the image of the nucleus, see Fig. \ref{fig: image charges}\color{red!50!black}a\color{black}. The first term is incorporated in the ionization energy (and the electron affinity respectively) in a vacuum, so our correction need not include this contribution.

Now consider a dielectric layer with nonzero thickness. We remove the real charges from the space, and insert infinite image charges on the right side such that the electric field felt by the electron is the same and certain boundary conditions on the left interface of the layer are satisfied: $E_{\parallelsum,outs.} = E_{\parallelsum,ins.}$ and $E_{\perp,outs.} = \varepsilon E_{\perp,ins.}$ where $ins.$ stands for the inside of the layer, $outs.$ for the outside of the layer, and $\varepsilon$ is the relative permittivity of the layer \cite{InfiniteImageCharges}. The infinite charges are depicted in Fig. \ref{fig: image charges}\color{red!50!black}b\color{black}. Each line of charges represents one of the image charges. The lines should coincide, but they are represented offset for clarity. Its charge values are given in the corresponding legend in the bottom left where we defined $\zeta := \frac{\varepsilon-1}{\varepsilon+1}$ for the sake of simplicity.

\enlargethispage{\baselineskip}

Let us define $S_{a}(z) := \sum_{k=0}^{\infty}\frac{z^{2k}}{(a+2k)^{2}}$. We calculate the force exerted on the electron by adding up the forces from the individual images:

\begin{gather}
\begin{aligned}
F &= \frac{-e}{4\pi\varepsilon_{0}}\Big{[}\frac{-\zeta q}{(x+2d)^{2}} + \frac{\zeta e}{(2x+2d)^{2}} + \frac{4\varepsilon}{(\varepsilon+1)^{2}}\Big{\{}\zeta \sum_{k=0}^{\infty}\frac{q\zeta^{2k}}{(x+2d+4h+4hk)^{2}}\hspace{1cm}\color{white}.\\
&\hspace{0.4cm}- \sum_{k=0}^{\infty}\frac{q\zeta^{2k}}{(x+2d+2h+4hk)^{2}}- \zeta \sum_{k=0}^{\infty}\frac{e\zeta^{2k}}{(2x+2d+4h+4hk)^{2}}\\
&\hspace{0.4cm}+ \sum_{k=0}^{\infty}\frac{e\zeta^{2k}}{(2x+2d+2h+4hk)^{2}}\Big{\}}\Big{]}
\end{aligned}
\label{eq: force}
\end{gather}

\begin{gather*}
\begin{aligned}
&= \frac{-e}{4\pi\varepsilon_{0}}\Big{[}\frac{\zeta q}{(x+2d)^{2}} + \frac{\zeta e}{(2x+2d)^{2}} + \frac{4\varepsilon}{(\varepsilon+1)^{2}(2h)^{2}}\Big{\{}\zeta qS_{\frac{x}{2h}+\frac{d}{h}+2}(\zeta) - qS_{\frac{x}{2h}+\frac{d}{h}+1}(\zeta)\\
&\hspace{1.1cm}- \zeta eS_{\frac{x}{h}+\frac{d}{h}+2}(\zeta) + eS_{\frac{x}{h}+\frac{d}{h}+1}(\zeta) \Big{\}}\Big{]}.
\end{aligned}
\end{gather*}

To find the potential, we integrate the expression in Eq. \ref{eq: force} from $x=0$ (at the nucleus) up to $x=\infty$ (the left direction is taken to be positive). We set $T_{a}(z) := \sum_{k=0}^{\infty}\frac{z^{2k}}{a+2k}$. Using the standard integrals i) $\int_{0}^{\infty}\frac{1}{(x+2d)^{2}}dx = \frac{1}{2d}$, ii) $\int_{0}^{\infty}\frac{1}{(2x+2d)^{2}}dx = \frac{1}{4d}$, and iii) $\int_{0}^{\infty}S_{ax+b}(z)dx =\\= \frac{1}{a}\sum_{k=0}^{\infty}\frac{z^{2k}}{b+2k} = \frac{1}{a}T_{b}(z)$, we obtain

\begin{gather}
\begin{aligned}
E &= \frac{-e}{4\pi\varepsilon_{0}}\Big{[}\frac{\zeta}{4d}(e-2q) + \frac{4\varepsilon}{(\varepsilon+1)^{2}(2h)^{2}}\Big{\{}2h\zeta qT_{\frac{d}{h}+2}(\zeta) - 2hqT_{\frac{d}{h}+1}(\zeta)\\
&\hspace{0.9cm}- h\zeta eT_{\frac{d}{h}+2}(\zeta) + heT_{\frac{d}{h}+1}(\zeta)\Big{\}}\Big{]}\\
&= \frac{-e}{4\pi\varepsilon_{0}}\Big{[}\frac{\zeta}{4d}(e-2q) + \frac{\varepsilon}{(\varepsilon+1)^{2}h}\big{\{}\zeta (2q-e)T_{\frac{d}{h}+2}(\zeta) - (2q-e)T_{\frac{d}{h}+1}(\zeta)\big{\}}\Big{]}\\
&= \frac{e(2q-e)}{4\pi\varepsilon_{0}}\Big{[}\frac{\zeta}{4d} + \frac{\varepsilon}{(\varepsilon+1)^{2}h}\big{\{}T_{\frac{d}{h}+1}(\zeta) - \zeta T_{\frac{d}{h}+2}(\zeta)\big{\}}\Big{]}.
\label{eq: energy}
\end{aligned}
\raisetag{100pt}
\end{gather}

\subsection{Parameter Space for Stability}

The potential correction in Eq. \ref{eq: energy} depends on the charge $q$ of the $\beta$-emitter, so we will write $E=E_{q}$. Henceforth, we will check the conditions per initial charge state. We summarize this for $Tm$ and $^{3}H$ as $\beta$-decayer in Table \ref{tab: Tm/Yb}. The first column indicates the initial charge state of the $\beta$-emitter, and each row lists the energy balance of ionization processes/electron captures. The values indicate the energy gained from the process, so stability requires all values in a single row to be negative.\footnote{The vacuum electron affinity of $Tm^{-}$ in the table is denoted by $E^{(vac.)}_{el.af.}$, as we do not have sufficient data available. We note that only the constraints from the other three columns are rather stringent.}

\begin{table*}[h]
\centering
\small
\scalebox{0.95}{
\begin{tabular}{p{2.2cm} | | l | l | l | l}
    \hline\hline
    \rule{0pt}{3.6ex}$Tm^{Q} \to Yb^{Q+1}$ & $Tm^{Q} \to Tm^{Q+1}$ & $Tm^{Q} \to Tm^{Q-1}$ & $Yb^{Q+1} \to Yb^{Q+2}$ & $Yb^{Q+1} \to Yb^{Q}$ \\ \hline\hline\rule{0pt}{2.5ex}
    $Tm^{-} \to Yb$ & -1.03 $eV + W + E_{0}$ & $E^{(vac.)}_{el.af.} - W - E_{-e}$ & -6.25 $eV + W + E_{e}$ & 0.02 $eV - W - E_{0}$ \\ \hline\rule{0pt}{2.5ex}
    $Tm \to Yb^{+}$ & -6.18 $eV + W + E_{e}$ & 1.03 $eV - W - E_{0}$ & -12.2 $eV + W + E_{2e}$ & 6.25 $eV - W - E_{e}$ \\ \hline\rule{0pt}{2.5ex}
    $Tm^{+} \to Yb^{2+}$ & -12.1 $eV + W + E_{2e}$ & 6.18 $eV - W - E_{e}$ & -25.1 $eV + W + E_{3e}$ & 12.2 $eV - W - E_{2e}$ \\ \hline\rule{0pt}{2.5ex}
    $Tm^{2+} \to Yb^{3+}$ & -23.7 $eV + W + E_{3e}$ & 12.1 $eV - W - E_{2e}$ & -43.6 $eV + W + E_{4e}$ & 25.1 $eV - W - E_{3e}$ \\
    \hline\hline
    \rule{0pt}{3.6ex} & $^{3}H \to \hspace{0.1cm}^{3}H^{+}$ & $^{3}H \to \hspace{0.1cm}^{3}H^{-}$ & $^{3}He^{+} \to\hspace{0.1cm}^{3}He^{2+}$ & $^{3}He^{+} \to\hspace{0.1cm}^{3}He$ \\ \hline\hline\rule{0pt}{2.5ex}
    $^{3}H \to\hspace{0.1cm}^{3}He^{+}$ & -13.60 $eV + W + E_{e}$ & 0.75 $eV - W - E_{0}$ & -54.4 $eV + W + E_{2e}$ & 0.5 $eV - W - E_{e}$ \\
\end{tabular}
}
\caption{The energy balance for electron transfer from/to the $\beta$-emitter \cite{Chemistry}.}
\label{tab: Tm/Yb}
\end{table*}

\enlargethispage{\baselineskip}

For any initial charge state of some other $\beta$-decayer candidates, such as $Ni$, $Eu$, $Sm$, and $Pu$, the columns $M^{Q} \to M^{Q+1}$ and $D^{Q+1} \to D^{Q}$ cannot be made simultaneously negative (where $M$ stands anew for mother and $D$ for daughter isotope). This is, however, possible for neutral $Pu$, but one would need to tune the parameters such that $6.03$ $eV$ $<E_{e}+W<6.27$ $eV$, which is not as stable as $Tm^{-}$. This means that these candidates permit no stable configuration with integer-charged states. Table \ref{tab: Tm/Yb} suggests $Tm^{-} \to Yb$ might be a stable configuration when $0.02$ $eV < E_{0}+W < 1.03$ $eV$. Furthermore, we want $W+E_{e} < 6.25$ $eV$ and $W + E_{-e} > E^{(vac.)}_{el.af.}$. Since we do not know $E^{(vac.)}_{el.af.}$, we omit the latter constraint and consider its effect later.

\begin{figure}[ht]
    \centering
    \begin{subfigure}[b]{0.49\textwidth}
        \centering
        \includegraphics[width=\textwidth]{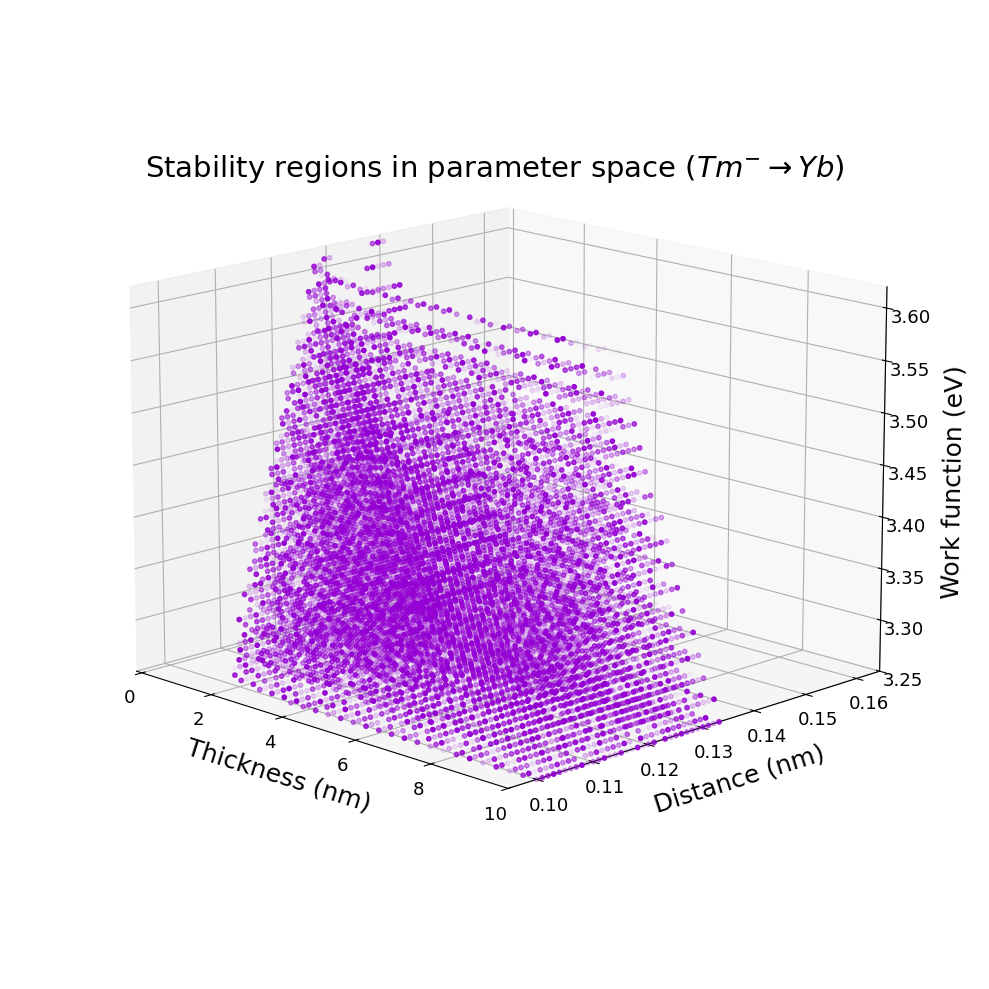}
        \caption{The stable configurations when the constraint, $E_{el.af.}^{(vac.)} - W - E_{-e} < 0$, is omitted.\\\color{white}{.}}
        \label{fig:stab1}
    \end{subfigure}
    \hfill
    \begin{subfigure}[b]{0.49\textwidth}
        \centering
        \includegraphics[width=\textwidth]{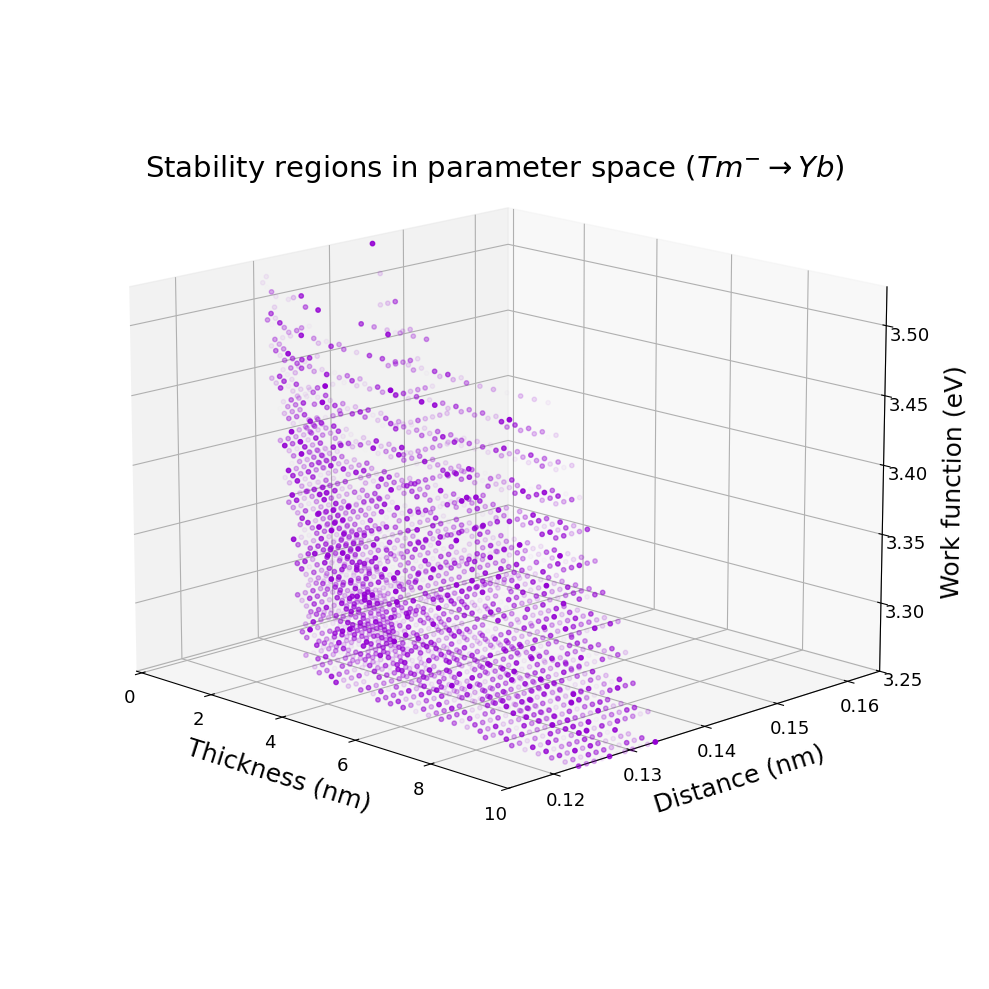}
        \caption{If $E_{el.af.}$ is $-4$ $eV$ and we account for the restriction, $E_{el.af.}^{(vac.)} - W - E_{-e} < 0$, we find fewer stable configurations.}
        \label{fig:stab2}
    \end{subfigure}
    \caption{The purple region indicates a stable configuration for both $Tm^{-}$ and $Yb$.}
    \label{fig: stability plots}
\end{figure}

To achieve these bounds, we tune the values of the distance between the $\beta$-emitter and the dielectric $d$, the thickness of the layer $h$, and the work function $W$. For each triplet $(d,h,W)$ that satisfies the constraints on $E_{0} + W$, we plot a purple dot in the parameter space, Fig. \ref{fig: stability plots}.

\vspace{-0.4cm}
\vspace{-0.2cm}

\begin{figure}
    \centering
    \includegraphics[width=0.53\textwidth]{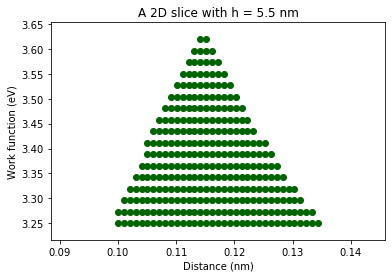}
    \caption{The green dots form a two-dimensional slice of Fig. \ref{fig:stab1} for a constant thickness $h$.}
    \label{fig: slice of stability plots}
\end{figure}

We see a purple, triangular-prism-shaped region. This means that the stability is rather independent of the thickness $h$. A slice of this region for constant thickness $h = 5.5$ $nm$ is represented in Fig. \ref{fig: slice of stability plots}.

It should be noted that the constraint $W + E_{-e} > E^{(vac.)}_{el.af.}$ is incorporated in Fig \ref{fig:stab2}, whereas it is omitted in Fig. \ref{fig:stab1} and \ref{fig: slice of stability plots}. For high values of $E^{(vac.)}_{el.af.}$, the purple region fades away and when it reaches $E^{(vac.)}_{el.af.} = -3$ $eV$ it completely disappears, unlike the same plot for $H \to He$, which is completely saturated with purple dots, i.e., the mother and daughter isotope are simultaneously stable. We therefore investigate the coupling between the $\beta$-emitter and the solid-state environment to allow for a nonvanishing $Tm^{Q} \to Yb^{Q}$ channel. This is prohibited for integer $Q$, so we treat the hybridized problem in Section \ref{sec: quantum}.

In the present analysis, the screening of the impurity potential by the graphene sheet has been described using the classical image-charge construction. This approximation is appropriate at sufficiently large distances, where the graphene layer behaves as a metallic conductor. At shorter length scales, however, the finite electronic compressibility of graphene leads to corrections associated with the quantum capacitance of the electron gas.

The validity of the classical approach can be estimated from the static susceptibility of graphene, $\chi_{0} \sim \frac{k_{F}}{\hbar v_{F}}$, where $k_{F} \sim \sqrt{\pi n}$ for an electron density $n$. Quantum corrections to the electrostatic response become relevant for length scales of order $(\frac{e^{2}}{\varepsilon}\chi_{0})^{-1}$. For typical electron densities, $n \sim 10^{12}$ cm$^{-2}$, this length is between $1$ and $10$ nm, which is comparable to the cumulative distances in Fig. \ref{fig:stab1} and \ref{fig:stab2}, i.e., the total distance between the $\beta$-emitter and the graphene sheet. This means that we cannot rigorously rule out the existence of stability regions when depositing the $\beta$-emitters on \textit{thin} dielectric spacer, as we did not account for the quantum corrections. It should be noted, however, that the goal of the dielectric slab is to maximize the distance between the radioactive atoms and the graphene sheet \cite{TanCheianov}.

\section{Turning on the Hybridization}
\label{sec: quantum}

We describe the solid-state environment of the $\beta$-decayers using scattering theory on a two-dimensional disk \cite{Anderson}. We assume the scattering states are $s$-waves, i.e., the azimuthal quantum number $l$ is equal to $0$. Then the single-particle wave functions are $\psi_{n}(r) = \frac{\sin\big{(}k_{n}r\big{)}}{\sqrt{\pi\ell r}}$, where the momenta $k_{n}$ are quantized, i.e., $k_{n} = \frac{2\pi n}{\ell}$ with $\ell$ the disk radius. In the presence of a local potential, the orbitals are distorted, an effect that is captured by a phase shift $\delta$. Specifically, the single-particle wave functions are $\varphi_{n}(r) = \frac{\sin\big{(}k_{n}r-\delta\cdot(1-\frac{r}{\ell})\big{)}}{\sqrt{\pi\ell r}}$.

The local potential is modified rapidly after the $\beta$-decay event, hence we are interested in the overlap between states in the absence of a local potential and states in the presence of a local potential. For a single-particle wave function, this is

\begin{equation}
\langle \psi_{n}\hspace{0.1cm}|\hspace{0.1cm}\varphi_{m}\rangle = 2\pi\int_{0}^{\ell}dr\hspace{0.1cm}r\hspace{0.1cm}\psi_{n}(r)\varphi_{m}(r) = \frac{\sin(\delta)}{\pi(m-n)+\delta},
\label{eq: single-particle state overlap}
\end{equation}

\noindent and for the full system, the amplitude is

\begin{equation}
\det\limits_{E_{n},E_{m}<E_{F}}\Big{(}\frac{\sin(\delta)}{\pi(m-n)+\delta}\Big{)}.
\label{eq: overlap absence/presence local potential}
\end{equation}

\subsection{Fermionic Model}

The different $\beta$-decay channels encode the spectral function. We elaborate on this function in Section \ref{subsec: spectral function}. The visibility of the $C\hspace{-0.05cm}\nu B$ spectrum can be protected if the spectral function has a power law IR divergence called the \textit{X-ray edge singularity}. For this reason, it is crucial to compute Eq. \ref{eq: overlap absence/presence local potential} for different local potentials corresponding to the various $\beta$-decay channels.

Assuming low orbital angular momenta reduces the problem's dimensionality. We claim that a one-dimensional fermionic system produces the same amplitude as in Eq. \ref{eq: overlap absence/presence local potential}. With the aid of a technique called \textit{bosonization} (see Section \ref{Bosonization}), we recast the Hamiltonian to a bosonic one, and in this language we proceed with the calculation of the spectral function.

Let $\psi = (R,L)^{T}$ be a massless one-dimensional Weyl fermion field, let $\sigma_{3} = \begin{psmallmatrix}
1 & 0\\
0 & -1
\end{psmallmatrix}$, and suppose the dynamics are governed by the Hamiltonian,

\begin{equation}
H_{0} = \int_{-\ell/2}^{\ell/2}dx\hspace{0.1cm}:\Big{(}\psi^{\dagger}[-iv_{F}\sigma_{3}\partial_{x}]\psi + V_{0}\delta(x)\psi^{\dagger}\psi\Big{)}:,
\label{eq: 1d massless Weyl fermions}
\end{equation}

\noindent for some local potential $V_{0}$, where $v_{F}$ is the Fermi velocity. Varying the action yields the wave equations:

\begin{equation}
\Big{[}\partial_{t} + v_{F}\partial_{x} + iV_{0}\delta(x) - \frac{V_{0}}{4\ell}\Big{]}L = 0 \hspace{0.4cm}\&\hspace{0.4cm}\Big{[}\partial_{t} - v_{F}\partial_{x} + iV_{0}\delta(x) - \frac{V_{0}}{4\ell}\Big{]}R = 0.
\end{equation}

Let us focus solely on the right-handed chirality sector ($R$) and analyze its solutions in the presence/absence of a local potential. The latter is obtained by setting $V_{0} = 0$. The quantized solutions to the wave equation are

\begin{equation}
R_{n}(x,t) = \frac{1}{\sqrt{\ell}}\exp\Big{(}i\big{[}k_{n}(x+v_{F}t) + \frac{V_{0}}{2v_{F}}\text{sgn}(x)\cdot(1-|\frac{x}{2\ell}|)\big{]}\Big{)}
\end{equation}

\noindent with $k_{n} = \frac{2\pi n}{\ell}$ and $n \in \mathbb{N}$. The (equal-time) overlap of $R_{m}$ and $R_{n}$ is thus

\begin{equation}
\begin{aligned}
\frac{1}{\ell}\int_{-\ell/2}^{\ell/2}dx\hspace{0.1cm}\exp\Big{(}\frac{2\pi i}{\ell}(n-m)x+i\frac{V_{0}}{2v_{F}}\text{sgn}(x)\cdot(1-|\frac{x}{2\ell}|)\Big{)} = \frac{\sin(\frac{V_{0}}{2v_{F}})}{\pi(m-n)+\frac{V_{0}}{2v_{F}}}.
\end{aligned}
\label{eq: duality 1d massless chiral Weyl fermions and low orbital momenta 2d scattering states}
\end{equation}

Eq. \ref{eq: duality 1d massless chiral Weyl fermions and low orbital momenta 2d scattering states} demonstrates the duality between our scattering problem and one-dimensional, massless Weyl fermions, via $\delta \leftrightarrow \frac{V_{0}}{2v_{F}}$. To capture the impurity states in this Hamiltonian, we supplement Eq. \ref{eq: 1d massless Weyl fermions} with the impurity field $d$ and $d^{\dagger}$, which introduces certain characteristic energy scales such as the on-site energy $E_{0}$, the Hubbard interaction $U$, and a conductance-impurity coupling $g$. Including furthermore an extrinsic spin variable $\sigma$, we summarize:

\begin{equation}
\begin{aligned}
H_{0} = \int dx\hspace{0.1cm}:\Big{(}&\psi_{\sigma}^{\dagger}[-iv_{F}\sigma_{3}\partial_{x}]\psi_{\sigma} + V_{0}\delta(x)\psi_{\sigma}^{\dagger}\psi_{\sigma} + g\delta(x)\psi_{\sigma}^{\dagger}\psi_{\sigma}d_{\sigma'}^{\dagger}d_{\sigma'}\\
&+ E_{0}\delta(x)(d_{\uparrow}^{\dagger}d_{\uparrow} + d_{\downarrow}^{\dagger}d_{\downarrow}) + U\delta(x)d_{\uparrow}^{\dagger}d_{\uparrow}d_{\downarrow}^{\dagger}d_{\downarrow}\Big{)}:.
\end{aligned}
\end{equation}

We focus anew solely on the right-handed chirality sector, and we incorporate an Anderson-type interaction term $H'$ between the conduction electrons and the impurity. Hence $H = H_{0} + H'$, with

\begin{equation}
\left\{\begin{array}{ll}
\begin{aligned}
\vspace{-0.2cm}H_{0} = \int dx\hspace{0.1cm}:\Big{(}&R_{\sigma}^{\dagger}[-iv_{F}\partial_{x}]R_{\sigma} + V_{0}\delta(x)R_{\sigma}^{\dagger}R_{\sigma} + g\delta(x)R_{\sigma}^{\dagger}R_{\sigma}d_{\sigma'}^{\dagger}d_{\sigma'}\\
&+ E_{0}\delta(x)d_{\sigma}^{\dagger}d_{\sigma} + U\delta(x)d_{\uparrow}^{\dagger}d_{\uparrow}d_{\downarrow}^{\dagger}d_{\downarrow}\Big{)}:,
\end{aligned}\vspace{0.3cm}\\
H' = \int dx\hspace{0.1cm}\delta(x)\hspace{0.1cm}\mathfrak{h}\hspace{0.05cm}R_{\sigma}^{\dagger}d_{\sigma} + h.c. = \mathfrak{h}\hspace{0.05cm}R_{\sigma}^{\dagger}(0)d_{\sigma} + h.c.
\end{array}\right.
\end{equation}

\noindent where $\mathfrak{h}$ is the hybridization parameter. Define:

\begin{subequations}
\begin{minipage}{0.45\textwidth}
\begin{flalign}
\hspace{-0.8cm}\left\{\begin{array}{ll}
\alpha = \frac{g}{2v_{F}}\int dx\hspace{0.05cm}:R_{\sigma}^{\dagger}(x)\text{ sgn}(x)R_{\sigma}(x):,\\
\beta = \frac{V_{0}}{2v_{F}}\int dx\hspace{0.05cm}:R_{\sigma}^{\dagger}(x)\text{ sgn}(x)R_{\sigma}(x):,
\end{array}\right.\hspace{-0.2cm}
\label{Alpha and beta}
\end{flalign}
\end{minipage}\hfill\begin{minipage}{0.51\textwidth}
\begin{flalign}
\vspace{-0.2cm}\color{black}\left\{\begin{array}{ll}
\Sigma(x) = \text{sgn}(x)\frac{V_{0}+gd_{\sigma}^{\dagger}d_{\sigma}}{2v_{F}},\\
S = \int dx\hspace{0.1cm}:R_{\sigma}^{\dagger}(x)\Sigma(x)R_{\sigma}(x):.
\end{array}\right.
\label{Sigma and S}
\end{flalign}
\end{minipage}
\label{Notation of objects for basis transformation}
\end{subequations}

Eq.s \ref{Alpha and beta} and \ref{Sigma and S} are related by $S = \alpha d_{\sigma}^{\dagger}d_{\sigma} + \beta$. The quantity $S$ can be used as a basis transformation to diagonalize the Hamiltonian $H_{0}$. Specifically, we map

\begin{equation}
\mathcal{O} \mapsto e^{iS}\mathcal{O}e^{-iS}.
\end{equation}

The $R$-fields, which we will refer to as the \textit{metallic degrees of freedom}, transform as $R_{\sigma}(x) \mapsto\\e^{-i\Sigma(x)}R_{\sigma}(x)$ and the $d$-fields transform as $d_{\sigma} \mapsto e^{-i\alpha}d_{\sigma}$. The latter follows from applying the identity $e^{izP} = 1 + (e^{iz}-1)P$ where $z \in \mathbb{C}$ and $P$ is an idempotent operator. As a result,

\begin{equation}
\left\{\begin{array}{ll}
H_{0} \mapsto \int dx\hspace{0.1cm}:\Big{(}R_{\sigma}^{\dagger}[-iv_{F}\partial_{x}]R_{\sigma} + E_{0}\delta(x)d_{\sigma}^{\dagger}d_{\sigma} + U\delta(x)d_{\uparrow}^{\dagger}d_{\uparrow}d_{\downarrow}^{\dagger}d_{\downarrow}\Big{)}:,\\
H' \mapsto \mathfrak{h}\hspace{0.05cm}R_{\sigma}^{\dagger}(0)e^{-i\alpha}d_{\sigma} + h.c.
\end{array}\right.
\end{equation}

The operators above are still in the Schrödinger picture. The spectral function can be expressed in terms of the hybridized initial and final states. Since we find those by adiabatically evolving the integer charge states with a unitary evolution operator depending on $H'$ in the \textit{interaction picture}, we perform the transformation:

\begin{equation}
\mathcal{O} \mapsto e^{iH_{0}t}\mathcal{O}e^{-iH_{0}t}.
\end{equation}

A time shift is generated by the Hamiltonian $H_{0}$, so e.g. $R_{\sigma}(x) = R_{\sigma}(x,0)$ is mapped to $R_{\sigma}(x,t)$. The $d$-fields also acquire a time dependence: $d_{\sigma} \mapsto e^{-itE_{0}}\big{(}1+(e^{-itU}-1)d_{\overline{\sigma}}^{\dagger}d_{\overline{\sigma}}\big{)}d_{\sigma}$ where $\overline{\sigma}$ is the opposite spin of $\sigma$. All in all,

\begin{equation}
\left\{\begin{array}{ll}
H_{0} = \int dx \Big{(}R_{\sigma}^{\dagger}[-iv_{F}\partial_{x}]R_{\sigma} +E_{0}\delta(x)d_{\sigma}^{\dagger}d_{\sigma} + U\delta(x)d_{\uparrow}^{\dagger}d_{\uparrow}d_{\downarrow}^{\dagger}d_{\downarrow}\Big{)},\\
H'(t) = \mathfrak{h}\hspace{0.05cm}R_{\sigma}^{\dagger}(0,t)e^{-\frac{ig}{2v_{F}}\int dx\hspace{0.1cm}:R_{\lambda}^{\dagger}(x,t)\text{ sgn}(x)\hspace{0.05cm}R_{\lambda}(x,t):}e^{-itE_{0}}\big{(}1+(e^{-itU}-1)d_{\overline{\sigma}}^{\dagger}d_{\overline{\sigma}}\big{)}d_{\sigma} + h.c.
\end{array}\right.
\label{Hamiltonians after basis transformation in the interaction picture}
\end{equation}

\noindent are the unperturbed Hamiltonian and the perturbation in the interaction picture. We find the unitary evolution operator,

\begin{equation}
U(-\infty,0) = \mathcal{T}\exp\Big{(}i\int_{-\infty}^{0}d\tau\hspace{0.1cm}e^{\varepsilon\tau}H'(\tau)\Big{)},
\label{Unitary evolution operator definition}
\end{equation}

\noindent where $\mathcal{T}$ denotes the time ordering symbol with $\varepsilon$ an infinitesimal, positive constant.

\subsection{Bosonization}
\label{Bosonization}

The metallic electrons have a dual bosonic description known as a \textit{Tomonaga-Luttinger liquid} (TLL). This means that the fermionic setting can be transformed to a bosonic one, albeit with unchanged local $d$-fields, simplifying our calculations of the correlation functions present in the spectral function. More specifically, we have two bosonic degrees of freedom, $\theta_{\sigma}$ and $\varphi_{\sigma}$, replacing $R_{\sigma}$ and $R_{\sigma}^{\dagger}$ via

\begin{equation}
\left\{\begin{array}{ll}
R_{\sigma}(x,t) = \frac{\kappa_{\sigma}^{-}}{\sqrt{2\pi a}}e^{-i\phi_{\sigma}(x,t)},\\
R_{\sigma}^{\dagger}(x,t) = \frac{\kappa_{\sigma}^{+}}{\sqrt{2\pi a}}e^{i\phi_{\sigma}(x,t)}
\end{array}\right.
\end{equation}

\noindent where $\phi_{\sigma} := \theta_{\sigma} + \varphi_{\sigma}$, where $a$ is a UV cut-off, and where the $\kappa_{\sigma}^{\pm}$ are the Klein factors, ensuring the correct fermionic commutation relations by $\kappa_{\uparrow(\downarrow)}^{\pm}\kappa_{\uparrow(\downarrow)}^{\mp} = 1$.

Suppressing the extrinsic spin variable in our notation for now, we express the bosonic commutation relations as

\begin{equation}
\left\{\begin{array}{ll}
\hspace{0cm}[\theta(x,t),\varphi(0,0)] = -\frac{i\pi}{4}\big{(}\text{sgn}(x-v_{F}t) + \text{sgn}(x+v_{F}t)\big{)},\\
\hspace{0cm}[\theta(x,t),\theta(0,0)] = \frac{i\pi}{4}\big{(}\text{sgn}(x-v_{F}t) - \text{sgn}(x+v_{F}t)\big{)} = [\varphi(x,t),\varphi(0,0)]
\end{array}\right.
\end{equation}

\noindent and the two-point correlators as

\begin{equation}
\langle\theta(x,t)\theta(0,0)\rangle = \frac{1}{2}\ln\big{(}\frac{x^{2}-v_{F}^{2}t^{2}+a^{2}-i\delta\text{ sgn}(t)}{\ell^{2}}\big{)} = \langle\varphi(x,t)\varphi(0,0)\rangle
\end{equation}

\noindent with $\delta$ an infinitesimal, positive constant selecting the principal branch \cite{Gogolin}. One can find the other two-point correlation functions using $\partial_{t}\varphi = v_{F}\partial_{x}\theta$. Some key correlators are summarized in Table \ref{Table: two-point correlators}.

\begin{table}[h]
\centering
\begin{tabular}{ |p{0.8cm}|p{0.8cm}||p{5.5cm}|p{2.75cm}|p{2.75cm}| }
 \hline
 \multicolumn{5}{|c|}{Limits of certain two-point correlators} \\
 \hline
 $x$ & $t$ & $\langle\theta(x,t)\theta(0,0)\rangle$ = $\langle\varphi(x,t)\varphi(0,0)\rangle$ & $\langle\varphi(x,t)\theta(0,0)\rangle$ & $\langle\theta(x,t)\varphi(0,0)\rangle$\\
 \hline
 $\neq 0$ & $=0$ & $\ln|\frac{x}{\ell}|$ & $-\frac{i\pi}{2}\text{sgn}(x)$ & $0$\\
 $=0$ & $\neq 0$ & $\ln|\frac{v_{F}t}{\ell}|+\frac{i\pi}{2}$ & $0$ & $\frac{i\pi}{2}$\\
 $=0$ & $=0$ & $\ln|\frac{a}{\ell}|$ & $0$ & $0$\\
 \hline
\end{tabular}
\caption{We assume that $t>0$ whenever $t\neq0$, as the correlators are time-ordered.}
\label{Table: two-point correlators}
\end{table}

Finally, we write down two normal ordered expressions in terms of bosons; particularly the ones in Eq. \ref{Hamiltonians after basis transformation in the interaction picture}. After the writing out the OPEs \cite{OPEs_bosonization}, one finds that

\begin{subequations}
\begin{align}
&:R_{\sigma}^{\dagger}R_{\sigma}:\hspace{0.2cm}\sim\hspace{0.2cm}\frac{1}{2\pi}\partial_{x}(\phi_{\uparrow} + \phi_{\downarrow}),\label{OPE1}\\
&:R_{\sigma}^{\dagger}\partial_{x}R_{\sigma}:\hspace{0.2cm}\sim\hspace{0.2cm}-\frac{i}{4\pi}(\partial_{x}\phi_{\uparrow})^{2} - \frac{i}{4\pi}(\partial_{x}\phi_{\downarrow})^{2}.\label{OPE2}
\end{align}
\label{OPEs}
\end{subequations}

\noindent Consequently,

\begin{equation}
\left\{\begin{array}{ll}
H_{0} = \int dx \Big{(}-\frac{v_{F}}{4\pi}(\partial_{x}\phi_{\uparrow})^{2}-\frac{v_{F}}{4\pi}(\partial_{x}\phi_{\downarrow})^{2} +E_{0}\delta(x)d_{\sigma}^{\dagger}d_{\sigma} + U\delta(x)d_{\uparrow}^{\dagger}d_{\uparrow}d_{\downarrow}^{\dagger}d_{\downarrow}\Big{)},\\
H'(t) = \frac{\mathfrak{h}\hspace{0.05cm}\kappa_{\sigma}^{+}}{\sqrt{2\pi a}}\hspace{0.05cm}e^{i\phi_{\sigma}(0,t)}e^{\frac{ig}{2\pi v_{F}}\big{(}\phi_{\uparrow}(0,t) + \phi_{\downarrow}(0,t)\big{)}}e^{-itE_{0}}\big{(}1+(e^{-itU}-1)d_{\overline{\sigma}}^{\dagger}d_{\overline{\sigma}}\big{)}d_{\sigma} + h.c.
\end{array}\right.
\label{Bosonized Hamiltonians}
\end{equation}

The perturbation $H'$ from Eq. \ref{Bosonized Hamiltonians} can be substituted into Eq. \ref{Unitary evolution operator definition}. We expand the time-ordered exponential up to first order in $\mathfrak{h}$ to obtain

\begin{equation}
\begin{aligned}
U(-\infty,0) &= \frac{1}{2}+\frac{i\mathfrak{h}}{\sqrt{2\pi a}}\int_{-\infty}^{0}\hspace{-0.3cm}dt\hspace{0.05cm}e^{\varepsilon t}\times\\
&\hspace{-0.3cm}\times\Big{(}\kappa_{\uparrow}^{+}e^{-i\phi_{\uparrow}(0,t)}e^{\frac{ig}{2\pi v_{F}}\big{(}\phi_{\uparrow}(0,t) + \phi_{\downarrow}(0,t)\big{)}} e^{-itE_{0}}\big{(}1+(e^{-itU}\hspace{-0.05cm}-1)d_{\downarrow}^{\dagger}d_{\downarrow}\big{)}d_{\uparrow}\\
&\hspace{-0.1cm}+\kappa_{\downarrow}^{+}e^{i\phi_{\downarrow}(0,t)}e^{\frac{ig}{2\pi v_{F}}\big{(}\phi_{\uparrow}(0,t) + \phi_{\downarrow}(0,t)\big{)}}e^{-itE_{0}}\big{(}1+(e^{-itU}\hspace{-0.05cm}-1)d_{\uparrow}^{\dagger}d_{\uparrow}\big{)}d_{\downarrow}\Big{)}+\hspace{-0.05cm}h.c. + \mathcal{O}(\mathfrak{h}^{2}).
\end{aligned}
\label{Unitary evolution operator in terms of bosonic fields}
\end{equation}

\subsection{X-ray Edge Singularity in the Spectral Function}
\label{subsec: spectral function}

At time $t=-\infty$, the initial state has an integer charge value, which we take as $|i(-\infty)\rangle :=\\|FS\rangle \otimes |00\rangle$. By applying $e^{iS}$ to this state, we map it to the same basis as the Hamiltonians in Eq. \ref{Bosonized Hamiltonians}. Hence, the hybridized initial state is $|i\rangle = U(-\infty,0)e^{iS}(|FS\rangle \otimes |00\rangle)$. We study the $\beta$-decay spectrum using the spectral function:

\begin{equation}
\mathcal{A}(E) := \sum_{\text{final states }f}|\langle i|f\rangle|^{2}\delta(E+E_{i}-E_{f}).
\label{Spectral function definition}
\end{equation}

We can write the delta function using the integral representation such that we integrate over the variable $t$. Since $H|f\rangle = E_{f}|f\rangle$, we can replace $e^{-iE_{f}t}$ by $e^{-iHt}$ if we let it act on the state $|f\rangle$. This results in

\begin{equation}
\begin{aligned}
\mathcal{A}(E) &= \sum_{f}|\langle i|f \rangle|^{2}\int\frac{dt}{2\pi}e^{i(E-E_{f}+E_{i})t}\\
&= \frac{1}{2\pi}\int dt\sum_{f}\langle i|e^{-iHt}|f\rangle\langle f|i\rangle e^{i(E+E_{i})t}.
\end{aligned}
\end{equation}

It follows that all contributions to the sum over final terms are independent of $f$, except for $|f\rangle\langle f|$, hence we can employ the identity $\sum\limits_{f}|f\rangle\langle f| = 1$. We find that the result is independent of the final state basis $\{|f\rangle\}$:

\begin{equation}
\mathcal{A}(E) = \frac{1}{2\pi}\int_{-\infty}^{\infty}dt\hspace{0.05cm}\langle i|e^{-iHt}|i\rangle e^{i(E+E_{i})t} =: \frac{1}{2\pi}\int_{-\infty}^{\infty}dt\hspace{0.05cm}C(t)e^{i(E+E_{i})t}
\end{equation}

\noindent where

\begin{equation}
C(t) = \underbrace{\overbrace{\Big{(}\langle 00| \otimes \langle FS|\Big{)}e^{-iS}}^{\substack{\color{red!50!black}\text{Writing the initial state}\\\color{red!50!black}\text{in the appropriate basis}}}U(0,-\infty)}_{\color{red!50!black}\text{Adiabatically turning on hybridization}}\exp\big{[}-iHt\big{]}\underbrace{U(-\infty,0)\overbrace{e^{iS}\Big{(}|FS\rangle \otimes |00\rangle\Big{)}}^{\substack{\color{red!50!black}\text{Writing the initial state}\\\color{red!50!black}\text{in the appropriate basis}}}}_{\color{red!50!black}\text{Adiabatically turning on hybridization}}.
\label{Correlation function C(t)}
\end{equation}

The zeroth order contribution describes the non-hybridized setting, and we have discussed the stability of the $\beta$-emitter in this setting with a classical argument in Section \ref{sec: E&M}. We therefore focus on higher orders in $\mathfrak{h}$.

First of all, we note that

\begin{equation}
e^{iS}\big{(}|FS\rangle \otimes |00\rangle\big{)} = e^{-\frac{iV_{0}}{2\pi v_{F}}\big{(}\phi_{\uparrow}(0,0) + \phi_{\downarrow}(0,0)\big{)}}\big{(}|FS\rangle \otimes |00\rangle\big{)},
\label{Basis transformation changes |FS>, but leave |00> invariant}
\end{equation}

\noindent since

\begin{equation}
\begin{aligned}
e^{iS} &= \Big{[}1 + (e^{-\frac{ig}{2\pi v_{F}}\big{(}\phi_{\uparrow}(0,0) + \phi_{\downarrow}(0,0)\big{)}}-1)(d_{\uparrow}^{\dagger}d_{\uparrow}+d_{\downarrow}^{\dagger}d_{\downarrow})\\
&\hspace{0.3cm}+(e^{-\frac{ig}{2\pi v_{F}}\big{(}\phi_{\uparrow}(0,0) + \phi_{\downarrow}(0,0)\big{)}}-1)^{2}d_{\uparrow}^{\dagger}d_{\uparrow}d_{\downarrow}^{\dagger}d_{\downarrow}\Big{]}e^{-\frac{iV_{0}}{2\pi v_{F}}\big{(}\phi_{\uparrow}(0,0) + \phi_{\downarrow}(0,0)\big{)}}.
\end{aligned}
\end{equation}

Secondly, we write out $\exp[-iHt] = \exp[-i(H_{0}+H')t]$ using the Dyson series expansion. The expanded part depending on $H'$ contains the relevant factors of $\mathfrak{h}$ and the part depending on $H_{0}$ can be written as a linear combination of $1$, $d_{\sigma}^{\dagger}d_{\sigma}$, and $d_{\uparrow}^{\dagger}d_{\uparrow}d_{\downarrow}^{\dagger}d_{\downarrow}$:

\begin{equation}
\begin{aligned}
e^{-iHt} &= \overbrace{\Lambda(t)\Big{(}1 + (e^{iE_{0}t}-1)d_{\sigma}^{\dagger}d_{\sigma} + (e^{iE_{0}t}-1)(e^{-i(E_{0}+U)t}-e^{-iUt}-2)d_{\uparrow}^{\dagger}d_{\uparrow}d_{\downarrow}^{\dagger}d_{\downarrow}\Big{)}}^{\color{red!50!black}\exp(-iH_{0}t)\color{black}}\times\\
&\hspace{0.3cm}\times\Big{(}1-i\int_{0}^{t}dt_{1}\hspace{0.05cm}H'(t_{1}) - \frac{1}{2}\int_{0}^{t}dt_{1}\int_{0}^{t}dt_{2}\hspace{0.05cm}\mathcal{T}\big{(}H'(t_{1})H'(t_{2})\big{)}\Big{)} + \mathcal{O}(\mathfrak{h}^{3})
\end{aligned}
\label{Expanding the exponentiated Hamiltonian using Dyson series}
\end{equation}

\noindent where $\Lambda(t) = \exp\Big{(}\frac{iv_{F}t}{4\pi}\int dx\hspace{0.05cm}\big{(}(\partial_{x}\phi_{\uparrow})^{2} + (\partial_{x}\phi_{\downarrow})^{2}\big{)}\Big{)}$.

One uses Eq. \ref{Unitary evolution operator in terms of bosonic fields} and Eq. \ref{Expanding the exponentiated Hamiltonian using Dyson series} to find that $C(t)$ has a vanishing first order term in $\mathfrak{h}$, as an odd number of $d$-fields are sandwiched between $|00\rangle$ and $\langle00|$. There are six second order terms:

\begin{enumerate}[label=\roman*)]
\setlength{\itemsep}{0pt}
    \item collect the second order term of $U(0,-\infty)$, and the zeroth order term of $\exp\big{[}-iHt\big{]}$ and $U(-\infty,0)$;
    \item collect the second order term of $\exp\big{[}-iHt\big{]}$, and the zeroth order term of $U(0,-\infty)$ and $U(-\infty,0)$;
    \item collect the second order term of $U(-\infty,0)$, and the zeroth order term of $\exp\big{[}-iHt\big{]}$ and $U(0,-\infty)$;
    \item collect the zeroth order term of $U(0,-\infty)$, and the first order term of $\exp\big{[}-iHt\big{]}$ and $U(-\infty,0)$;
    \item[\textbf{v)}] collect the \textbf{zeroth} order term of $\underline{\exp\big{[}-iHt\big{]}}$, and the \textbf{first} order term of $\underline{U(0,-\infty)}$ and $\underline{U(-\infty,0)}$;
    \item[vi)] collect the zeroth order term of $U(-\infty,0)$, and the first order term of $\exp\big{[}-iHt\big{]}$ and $U(0,-\infty)$.
\end{enumerate}

The fifth contribution is dominant, for the first three are virtual processes that stay finite in the large $t$ limit, and the fourth and sixth term vanish in the large $t$ limit. Henceforth, we only account for the fifth term for this reason.

The terms in Eq. \ref{Unitary evolution operator in terms of bosonic fields} with three $d$-fields vanish in $C(t)$, for they annihilate $|00\rangle$. A physical interpretation of this statement is that this process is an electron hopping to the impurity and back, which does not involve any double occupation states with the additional energy $U$. Similarly, the terms in $C(t)$ with three $d$-fields from $U(0,-\infty)$ vanish, as these fields annihilate $\langle00|$. Finally, the $d_{\uparrow}^{\dagger}d_{\uparrow}d_{\downarrow}^{\dagger}d_{\downarrow}$-term in Eq. \ref{Expanding the exponentiated Hamiltonian using Dyson series} does not contribute to the second order of $\mathfrak{h}$ in $C(t)$ for the same reason. All in all, if we neglect the zeroth order contribution and higher than second order contributions, then we see that

\begin{equation}
\begin{aligned}
C(t) &= -\frac{\mathfrak{h}^{2}}{2\pi a}e^{-iE_{0}t}\int_{-\infty}^{0}dt_{1}\int_{-\infty}^{0}dt_{2}\hspace{0.05cm}e^{\varepsilon(t_{1}+t_{2})}\Big{(}\langle00|\otimes\langle FS|\Big{)}e^{\frac{iV_{0}}{2\pi v_{F}}\Big{(}\phi_{\uparrow}(0,0) + \phi_{\downarrow}(0,0)\Big{)}}\times\\
&\hspace{-0.9cm}\times\Big{(}e^{-i\phi_{\uparrow}(0,t_{1})}e^{\frac{ig}{2\pi v_{F}}\Big{(}\phi_{\uparrow}(0,t_{1})+\phi_{\downarrow}(0,t_{1})\Big{)}}e^{-it_{1}E_{0}}d_{\uparrow}\Lambda(t)d_{\uparrow}^{\dagger}e^{it_{2}E_{0}}e^{-\frac{ig}{2\pi v_{F}}\Big{(}\phi_{\uparrow}(0,t_{2})+\phi_{\downarrow}(0,t_{2})\Big{)}}e^{i\phi_{\uparrow}(0,t_{2})}\\
&\hspace{-0.7cm}+ e^{-i\phi_{\downarrow}(0,t_{1})}e^{\frac{ig}{2\pi v_{F}}\Big{(}\phi_{\uparrow}(0,t_{1})+\phi_{\downarrow}(0,t_{1})\Big{)}}e^{-it_{1}E_{0}}d_{\downarrow}\Lambda(t)d_{\downarrow}^{\dagger}e^{it_{2}E_{0}}e^{-\frac{ig}{2\pi v_{F}}\Big{(}\phi_{\uparrow}(0,t_{2})+\phi_{\downarrow}(0,t_{2})\Big{)}}e^{i\phi_{\uparrow}(0,t_{2})}\Big{)}\times\\
&\hspace{-0.9cm}\times e^{-\frac{iV_{0}}{2\pi v_{F}}\Big{(}\phi_{\uparrow}(0,0) + \phi_{\downarrow}(0,0)\Big{)}}\Big{(}|FS\rangle\otimes|00\rangle\Big{)}\\
&\\
&= -\frac{\mathfrak{h}^{2}}{\pi a}e^{iE_{0}t}\int_{-\infty}^{0}dt_{1}\int_{-\infty}^{0}dt_{2}\hspace{0.05cm}e^{\varepsilon(t_{1}+t_{2})}e^{-i(t_{1}-t_{2})E_{0}}\times\\
&\hspace{-0.9cm}\times\langle FS|e^{\frac{iV_{0}}{2\pi v_{F}}\phi(0,0)}e^{-i(1-\frac{g}{2\pi v_{F}})\phi(0,t_{1})}\Lambda(t)e^{i(1-\frac{g}{2\pi v_{F}})\phi(0,t_{2})}e^{-\frac{iV_{0}}{2\pi v_{F}}\phi(0,0)}|FS\rangle\times\\
&\hspace{-0.9cm}\times\langle FS|e^{\frac{iV_{0}}{2\pi v_{F}}\phi(0,0)}e^{\frac{ig}{2\pi v_{F}}\phi(0,t_{1})}\Lambda(t)e^{-\frac{ig}{2\pi v_{F}}\phi(0,t_{2})}e^{-\frac{iV_{0}}{2\pi v_{F}}\phi(0,0)}|FS\rangle\\
&\\
&= -\frac{\mathfrak{h}^{2}}{\pi a}e^{i(E_{0}-E_{FS})t}\int_{-\infty}^{0}dt_{1}\int_{-\infty}^{0}dt_{2}\hspace{0.05cm}e^{\varepsilon(t_{1}+t_{2})}e^{-i(t_{1}-t_{2})E_{0}}\times\\
&\hspace{-0.9cm}\times\langle FS|e^{\frac{iV_{0}}{2\pi v_{F}}\phi(0,0)}e^{-i(1-\frac{g}{2\pi v_{F}})\phi(0,t_{1})}e^{i(1-\frac{g}{2\pi v_{F}})\phi(0,t+t_{2})}e^{-\frac{iV_{0}}{2\pi v_{F}}\phi(0,t)}|FS\rangle\times\\
&\hspace{-0.9cm}\times\langle FS|e^{\frac{iV_{0}}{2\pi v_{F}}\phi(0,0)}e^{\frac{ig}{2\pi v_{F}}\phi(0,t_{1})}e^{-\frac{ig}{2\pi v_{F}}\phi(0,t+t_{2})}e^{-\frac{iV_{0}}{2\pi v_{F}}\phi(0,t)}|FS\rangle.
\end{aligned}
\label{Writing C(t) as the product of two correlators}
\end{equation}

The second equality exploits the symmetry of the extrinsic spin: interchanging $\uparrow$ and $\downarrow$ leaves $C(t)$ invariant. The third equality is based on the time translation by $t$ due to the operator $\Lambda(t)$. The result depends on the product of two correlators. Both correlators take the form,

\begin{equation}
\Pi_{A}(t,t_{1},t_{2}) := \langle FS|e^{\frac{iV_{0}}{2\pi v_{F}}\phi(0,0)}e^{iA\phi(0,t_{1})}e^{-iA\phi(0,t+t_{2})}e^{-\frac{iV_{0}}{2\pi v_{F}}\phi(0,t)}|FS\rangle,
\end{equation}

\noindent where $A = \frac{g}{2\pi v_{F}} - 1$ for the first correlators in the product and $A = \frac{g}{2\pi v_{F}}$ for the second correlator. We can write

\begin{equation}
\Pi_{A}(t,t_{1},t_{2}) = f_{A}(t,t_{1},t_{2})\langle FS|\mathcal{T}\Big{(}e^{\frac{iV_{0}}{2\pi v_{F}}\phi(0,0)}e^{iA\phi(0,t_{1})}e^{-iA\phi(0,t+t_{2})}e^{-\frac{iV_{0}}{2\pi v_{F}}\phi(0,t)}\Big{)}|FS\rangle,
\label{Time ordered Pi correlation}
\end{equation}

\noindent where

\begin{equation}
\begin{aligned}
f_{A}(t,t_{1},t_{2}) &= \Big{(}e^{-i\frac{AV_{0}}{2v_{F}}}\omega(t_{1}-t) + e^{i\pi(\frac{V_{0}}{2\pi v_{F}})^{2}}\omega(t)\omega(t_{1}-t_{2}-t)\\
&\hspace{0.3cm}+ \omega(t)\omega(t+t_{2}-t_{1})\omega(-t_{2}-t) + e^{i\frac{AV_{0}}{2v_{F}}}\omega(t)\omega(t_{2}+t)\\
&\hspace{0.3cm}+\omega(t-t_{1})\omega(-t)\omega(t_{1}-t_{2}-t)+e^{i\pi A^{2}}\omega(t_{2}+t-t_{1})\omega(-t)\Big{)}
\end{aligned}
\end{equation}

\noindent with $\omega$ the Heaviside step function using the BCH formula persistently. The derivation can be found in Appendix \ref{Artificial Time ordering}.

The Gaussianity of the field $\phi$ tells us that the cumulant expansion of Eq. \ref{Time ordered Pi correlation} terminates after the second order. This means that

\begin{equation}
\langle e^{W}e^{X}e^{Y}e^{Z}\rangle = e^{\frac{1}{2}\big{(}\langle W^{2}\rangle + \langle X^{2}\rangle + \langle Y^{2}\rangle + \langle Z^{2}\rangle\big{)}}e^{\langle WX\rangle + \langle WY\rangle + \langle WZ\rangle + \langle XY\rangle + \langle XZ\rangle + \langle YZ\rangle}
\end{equation}

\noindent where $\langle\cdot\rangle := \langle FS|\mathcal{T}(\cdot)|FS\rangle$ and where $W$, $X$, $Y$, $Z$ are of the form $\lambda\phi(x,t)$ for some $\lambda \in \mathbb{C}$. Using the two-point correlators in Table \ref{Table: two-point correlators} and $\phi = \theta + \varphi$, we compute

\begin{equation}
\begin{aligned}
\Pi_{A}(t,t_{1},t_{2}) &= f_{A}(t,t_{1},t_{2})\overbrace{e^{\frac{1}{2}\big{(}-2(\frac{V_{0}}{2\pi v_{F}})^{2}\cdot2\ln|\frac{a}{\ell}|-2A^{2}\cdot2\ln|\frac{a}{\ell}|\big{)}}}^{\color{red!50!black}e^{\frac{1}{2}\big{(}\langle W^{2}\rangle+\langle X^{2}\rangle + \langle Y^{2}\rangle +\langle Z^{2}\rangle\big{)}}\color{black}}\cdot e^{-\frac{AV_{0}}{2\pi v_{F}}\ln|\frac{v_{F}t_{1}}{\ell}|+\frac{AV_{0}}{2\pi v_{F}}\ln|\frac{v_{F}(t+t_{2})}{\ell}|}\times\\
&\hspace{0.3cm}\times e^{(\frac{V_{0}}{2\pi v_{F}})^{2}\ln|\frac{v_{F}t}{\ell}|+A^{2}\ln|\frac{v_{F}(t+t_{2}-t_{1})}{\ell}|+\frac{AV_{0}}{2\pi v_{F}}\ln|\frac{v_{F}(t-t_{1})}{\ell}|-\frac{AV_{0}}{2\pi v_{F}}\ln|\frac{v_{F}t_{2}}{\ell}|}\cdot e^{\frac{3\pi i}{2}\big{(}A^{2}+(\frac{V_{0}}{2\pi v_{F}})^{2}\big{)}}\\
&= f_{A}(t,t_{1},t_{2})e^{\frac{3\pi i}{2}\big{(}A^{2}+(\frac{V_{0}}{2\pi v_{F}})^{2}\big{)}}\big{|}\frac{a}{\ell}\big{|}^{-2\big{(}A^{2}+(\frac{V_{0}}{2\pi v_{F}})^{2}\big{)}}\big{|}\frac{v_{F}t}{\ell}\big{|}^{(\frac{V_{0}}{2\pi v_{F}})^{2}}\big{|}\frac{v_{F}t_{1}}{\ell}\big{|}^{-\frac{AV_{0}}{2\pi v_{F}}}\times\\
&\hspace{0.3cm}\times\big{|}\frac{v_{F}t_{2}}{\ell}\big{|}^{-\frac{AV_{0}}{2\pi v_{F}}}\big{|}\frac{v_{F}(t-t_{1})}{\ell}\big{|}^{\frac{AV_{0}}{2\pi v_{F}}}\big{|}\frac{v_{F}(t+t_{2})}{\ell}\big{|}^{\frac{AV_{0}}{2\pi v_{F}}}\big{|}\frac{v_{F}(t+t_{2}-t_{1})}{\ell}\big{|}^{A^{2}},
\end{aligned}
\end{equation}

\noindent thus

\begin{equation}
\begin{aligned}
\mathcal{A}(E) &= -\frac{\mathfrak{h}^{2}}{2\pi^{2}a}e^{(\frac{3\pi i}{2}-2\ln|\frac{a}{\ell}|)\big{(}1-\frac{g}{\pi v_{F}}+(\frac{V_{0}}{2\pi v_{F}})^{2}\big{)}}\int_{-\infty}^{\infty}dt\int_{-\infty}^{0}dt_{1}\int_{-\infty}^{0}dt_{2}\hspace{0.1cm}e^{i(E+E_{i}+E_{0}-E_{FS})t}\times\\
&\hspace{0.3cm}\times e^{\varepsilon(t_{1}+t_{2})}e^{-iE_{0}(t_{1}-t_{2})}f_{\frac{g}{2\pi v_{F}}-1}(t,t_{1},t_{2})f_{\frac{g}{2\pi v_{F}}}(t,t_{1},t_{2})\times\\
&\hspace{0.3cm}\times\big{|}\frac{v_{F}t}{\ell}\big{|}^{N_{1}}\big{|}\frac{v_{F}t_{1}}{\ell}\big{|}^{N_{2}}\big{|}\frac{v_{F}t_{2}}{\ell}\big{|}^{N_{2}}\big{|}\frac{v_{F}(t-t_{1})}{\ell}\big{|}^{-N_{2}}\big{|}\frac{v_{F}(t+t_{2})}{\ell}\big{|}^{-N_{2}}\big{|}\frac{v_{F}(t+t_{2}-t_{1})}{\ell}\big{|}^{N_{3}},
\end{aligned}
\label{Spectral function as a volume integral}
\end{equation}

\noindent where

\begin{equation}
    \begin{aligned}
    \left\{\begin{array}{ll}
    N_{1} = 2(\frac{V_{0}}{2\pi v_{F}})^{2},\\
    N_{2} = \frac{V_{0}}{2\pi v_{F}} - \frac{gV_{0}}{2\pi^{2} v_{F}^{2}},\\
    N_{3} = 1 - \frac{g}{\pi v_{F}} + 2(\frac{g}{2\pi v_{F}})^{2}.
    \end{array}\right.
    \end{aligned}
\end{equation}

Recall that we aim to explore the region in the parameter space for which the spectral function has an IR divergence \cite{Infinite_hole_mass,Many_electron_singularity,Nozieres_I,Nozieres_II,Nozieres_III,SchotteSchotte}. This boils down to the integrability of the volume integral in Eq. \ref{Spectral function as a volume integral}. We know that $N_{1} > -1$ and $N_{3} > -1$ (as $|g| \ll v_{F}$), so we consider two cases: (i) $N_{2} < -1$ and (ii) $N_{2} > -1$.

\subsubsection*{$\mathbf{N_{2} < -1}$}

The $t_{1}$- and $t_{2}$-integral have non-integrable integrands with the divergences occurring at $t_{1} \to 0$ and $t_{2} \to 0$ (and not at any other value). The dominant contribution of both integrals is at $0$, so we replace every $t_{1}$ and $t_{2}$ with $0$ with an accuracy of $\frac{a}{v_{F}}$, i.e., we also replace $|\frac{v_{F}t_{1}}{\ell}|$ and $|\frac{v_{F}t_{2}}{\ell}|$ with $|\frac{a}{\ell}|$:

\begin{equation}
\begin{aligned}
\mathcal{A}(E) &= -\frac{\mathfrak{h}^{2}}{2\pi^{2}a}e^{(\frac{3\pi i}{2}-2\ln|\frac{a}{\ell}|)\big{(}1-\frac{g}{\pi v_{F}}+(\frac{V_{0}}{2\pi v_{F}})^{2}\big{)}}\big{|}\frac{a}{\ell}\big{|}^{2N_{2}}\int_{-\infty}^{\infty}dt\hspace{0.1cm}e^{i(E+E_{i}+E_{0}-E_{FS})t}\times\\
&\hspace{0.3cm}\times f_{\frac{g}{2\pi v_{F}}-1}(t,\color{red!50!black}0\color{black},\color{red!50!black}0\color{black})f_{\frac{g}{2\pi v_{F}}}(t,\color{red!50!black}0\color{black},\color{red!50!black}0\color{black})\big{|}\frac{v_{F}t}{\ell}\big{|}^{N_{1}}\color{red!50!black}\big{|}\frac{v_{F}t}{\ell}\big{|}^{-N_{2}}\big{|}\frac{v_{F}t}{\ell}\big{|}^{-N_{2}}\big{|}\frac{v_{F}t}{\ell}\big{|}^{N_{3}}\color{black}\\
&\propto |E+E_{i}+E_{0}-E_{FS}|^{-(N_{1}-2N_{2}+N_{3}+1)}.
\end{aligned}
\end{equation}

We see that the X-ray edge singularity is present for $N_{2} < -1$. The last step is based on the definition of the Gamma function, as $f_{A}(x,y,z)$ is piecewise constant. This relies on the fact that $N_{1} - 2N_{2} + N_{3} > -1$, which follows from

\begin{equation}
N_{1} - 2N_{2} + N_{3} = \frac{1}{2}\Big{(}(\frac{g}{\pi v_{F}}+\frac{V_{0}}{2\pi v_{F}} - 1)^{2} + \frac{3V_{0}^{2}}{4\pi^{2}v_{F}^{2}} + 1\Big{)} > 0 > -1.
\label{Eq: x-ray edge exponent when N_2 < -1}
\end{equation}

\subsubsection*{$\mathbf{N_{2} > -1}$}

The non-integrable singularities now lie at $t_{1} \to t$ and $t_{2} \to -t$. We perform the $t_{2}$-integral first, retaining only the dominant contribution of this integral and changing the bound of the $t$-integral for consistency:

\begin{equation}
\begin{aligned}
\mathcal{A}(E) &= -\frac{\mathfrak{h}^{2}}{2\pi^{2}a}e^{(\frac{3\pi i}{2}-2\ln|\frac{a}{\ell}|)\big{(}1-\frac{g}{\pi v_{F}}+(\frac{V_{0}}{2\pi v_{F}})^{2}\big{)}}\int_{0}^{\infty}\hspace{-0.3cm}dt\int_{-\infty}^{0}\hspace{-0.3cm}dt_{1}\hspace{0.1cm}e^{i(E+E_{i}+E_{0}-E_{FS})t}\times\\
&\hspace{0.3cm}\times e^{\varepsilon(t_{1}\color{red!50!black}-t\color{black})}e^{-iE_{0}(t_{1}\color{red!50!black}+t\color{black})}f_{\frac{g}{2\pi v_{F}}-1}(t,t_{1},\color{red!50!black}-t\color{black})f_{\frac{g}{2\pi v_{F}}}(t,t_{1},\color{red!50!black}-t\color{black})\big{|}\frac{v_{F}t}{\ell}\big{|}^{N_{1}}\big{|}\frac{v_{F}t_{1}}{\ell}\big{|}^{N_{2}}\color{red!50!black}\big{|}\frac{v_{F}t}{\ell}\big{|}^{N_{2}}\color{black}\times\\
&\hspace{0.3cm}\times\big{|}\frac{v_{F}(t-t_{1})}{\ell}\big{|}^{-N_{2}}\big{|}\frac{a}{\ell}\big{|}^{-N_{2}}\color{red!50!black}\big{|}\frac{v_{F}t_{1}}{\ell}\big{|}^{N_{3}}\color{black}\\
&=-\frac{\mathfrak{h}^{2}}{2\pi^{2}a}e^{(\frac{3\pi i}{2}-2\ln|\frac{a}{\ell}|)\big{(}1-\frac{g}{\pi v_{F}}+(\frac{V_{0}}{2\pi v_{F}})^{2}\big{)}}\big{|}\frac{a}{\ell}\big{|}^{-N_{2}}\int_{0}^{\infty}\hspace{-0.3cm}dt\int_{-\infty}^{0}\hspace{-0.3cm}dt_{1}\hspace{0.1cm}e^{i(E+E_{i}-E_{FS})t}e^{\varepsilon(t_{1}-t)}\times\\
&\hspace{0.3cm}\times e^{-iE_{0}t_{1}}f_{\frac{g}{2\pi v_{F}}-1}(t,t_{1},-t)f_{\frac{g}{2\pi v_{F}}}(t,t_{1},-t)\big{|}\frac{v_{F}t}{\ell}\big{|}^{N_{1}+N_{2}}\big{|}\frac{v_{F}t_{1}}{\ell}\big{|}^{N_{2}+N_{3}}\big{|}\frac{v_{F}(t-t_{1})}{\ell}\big{|}^{-N_{2}}.
\end{aligned}
\end{equation}

The remaining area integral has a non-integrable divergence at $t_{1} \to t$, but this occurs only when $(t,t_{1}) \to (0,0)$. We study this point with the coordinate transformation

\begin{equation}
\left\{\begin{array}{ll}
u = t + t_{1}\\
v = t - t_{1}
\end{array}\right.\hspace{1cm}\implies\hspace{1cm}\big{|}\det\big{(}\mathbf{J}(u,v)\big{)}\big{|} = \frac{1}{2}.
\end{equation}

The integration bounds change accordingly. The integral diverges as $v\to0$, so we keep only its dominant contribution once again:

\begin{equation}
\begin{aligned}
\mathcal{A}(E) &= -\frac{\mathfrak{h}^{2}}{4\pi^{2}a}e^{(\frac{3\pi i}{2}-2\ln|\frac{a}{\ell}|)\big{(}1-\frac{g}{\pi v_{F}}+(\frac{V_{0}}{2\pi v_{F}})^{2}\big{)}}\big{|}\frac{a}{\ell}\big{|}^{-N_{2}}\int_{-\infty}^{0}\hspace{-0.3cm}du\int_{u}^{-u}\hspace{-0.3cm}dv\hspace{0.1cm}e^{i(E+E_{i}-E_{FS})\frac{u+v}{2}}e^{-\varepsilon v}\times\\
&\hspace{0.3cm}\times e^{-iE_{0}\frac{u-v}{2}}f_{\frac{g}{2\pi v_{F}}-1}(\frac{u+v}{2},\frac{u-v}{2},-\frac{u+v}{2})f_{\frac{g}{2\pi v_{F}}}(\frac{u+v}{2},\frac{u-v}{2},-\frac{u+v}{2})\times\\
&\hspace{0.3cm}\times\big{|}\frac{v_{F}(u+v)}{2\ell}\big{|}^{N_{1}+N_{2}}\big{|}\frac{v_{F}(u-v)}{2\ell}\big{|}^{N_{2}+N_{3}}\big{|}\frac{v_{F}v}{\ell}\big{|}^{-N_{2}}
\end{aligned}
\end{equation}

\begin{equation*}
\begin{aligned}
&= -\frac{\mathfrak{h}^{2}}{4\pi^{2}a}e^{(\frac{3\pi i}{2}-2\ln|\frac{a}{\ell}|)\big{(}1-\frac{g}{\pi v_{F}}+(\frac{V_{0}}{2\pi v_{F}})^{2}\big{)}}\big{|}\frac{a}{\ell}\big{|}^{-2N_{2}}\int_{-\infty}^{0}\hspace{-0.3cm}du\hspace{0.1cm}e^{i(E+E_{i}-E_{FS})\frac{u}{2}}\times\\
&\hspace{0.3cm}\times e^{-iE_{0}\frac{u}{2}}f_{\frac{g}{2\pi v_{F}}-1}(\frac{u}{2},\frac{u}{2},-\frac{u}{2})f_{\frac{g}{2\pi v_{F}}}(\frac{u}{2},\frac{u}{2},-\frac{u}{2})\big{|}\frac{v_{F}u}{2\ell}\big{|}^{N_{1}+2N_{2}+N_{3}}\\
&\propto |E+E_{i}-E_{0}-E_{FS}|^{-(N_{1}+2N_{2}+N_{3}+1)}.
\end{aligned}
\end{equation*}

We see that the X-ray edge singularity is present for $N_{2} > -1$ too. In the final step, we used the definition of the Gamma function once again. This is allowed, for

\begin{equation}
N_{1} + 2N_{2} + N_{3} = \frac{1}{2}\Big{(}(\frac{g}{\pi v_{F}} - \frac{V_{0}}{\pi v_{F}} - 1)^{2} + 1\Big{)} > 0 > -1.
\label{Eq: x-ray edge exponent when N_2 > -1}
\end{equation}

In the conventional Mahan–Nozières–De Dominicis description of the X-edge singularity the threshold exponent contains two contributions: a linear term in the scattering phase shift arising from excitonic correlations between the excited electron and the core hole, and a quadratic term associated with Anderson’s orthogonality catastrophe. In light of this theory, the exponent is proportional to $-\frac{2\delta}{\pi} + (\frac{\delta}{\pi})^{2}$ \cite{Nozieres_I,Nozieres_II,Nozieres_III}.

In the present $\beta$-decay problem the emitted electron is created by the weak interaction rather than promoted from the Fermi sea. For this reason we cannot expect the same ratio between the contributions by the excitonic term and the Anderson term to the singularity. We can insert the phase shift $\delta = \frac{V_{0}}{2v_{F}}$ in Eq. \ref{Eq: x-ray edge exponent when N_2 < -1} and Eq. \ref{Eq: x-ray edge exponent when N_2 > -1} to see that both the linear and the quadratic terms are still present, albeit with different coefficients.

The monomial dependence on the energy introduces no length scale, and this description breaks down near the spectral edge. Therefore, we multiply our spectral function by a factor of $\exp\big({-\frac{Ed}{\hbar v_{F}}}\big)$, where $d$ is the distance between the substrate and the Coulomb potential. The resulting function is a Gamma distribution, and one can estimate the linewidth $\xi$ as the absolute value of its first moment. This is the quotient of two Gamma functions, yielding $\frac{\hbar v_{F}}{d}(1-\alpha_{\pm})$, where we define $\alpha_{\pm} := N_{1} \pm 2N_{2} + N_{3} + 1$. The approximate linewidth is therefore $\frac{\hbar v_{F}}{d}(\alpha_{\pm} - 1)$, since $\alpha_{\pm} > 1$. This width is suppressed for values of $\alpha_{\pm}$ close to $1$. The smallest possible value of $\alpha_{\pm}$ is $\frac{3}{2}$ (for instance, when $V_{0} \ll v_{F}$ and $g \sim \pi v_{F}$), which yields $\xi \sim \frac{\hbar v_{F}}{2d}$. For graphene, this means $\xi \sim 3$ $eV$ for $d = 0.1$ nm, $\xi \sim 0.3$ $eV$ for $d = 1$ nm, etc.

A more precise estimate would require convolving the bare spectrum with the spectral function, but this is beyond the scope of this paper. We remark that the large Fermi velocity of graphene enhances broadening, so one might consider alternative substrates, such as pressurized $\alpha$-(BEDT-TTF)$_{2}I_{3}$ \cite{alphaBEDTTTF2I3} or $1T$-$CrO_{2}$ monolayers \cite{1TCrO2}.

\section{Conclusion and Outlook}

Inserting a dielectric slab between the $\beta$-decayer and the solid-state environment introduces Coulomb interactions, causing the disappearance of stability regions in the parameter space plots in Fig. \ref{fig: stability plots}, as shown in Section \ref{sec: E&M}. This is detrimental to the visibility of the $C\hspace{-0.05cm}\nu B$-spectrum.

Rather than increasing the distance between the $\beta$-emitter via a dielectric spacer, an alternative solution is suggested for which the impurities are directly deposited onto a substrate with a high dielectric constant may enhance the $C\hspace{-0.05cm}\nu B$-visibility \cite{TanCheianov}. In Section \ref{sec: quantum}, we included the effect of the substrate-impurity coupling up to second order and found an X-ray edge singularity, which could potentially remedy the visibility. We assumed a weak but nonzero coupling, and we expect the presence of a singularity to persist for strong coupling as well. We hope this supposition will be investigated numerically in future research activities.

More importantly, we expect our findings to be useful for experiments involving relic neutrino detection. We note, however, that it is vital to account for more quantum mechanical effects to optimize for an experiment. We provide a non-exhaustive list of such effects to be examined in further research:

\begin{enumerate}[label=\roman*)]
\setlength{\itemsep}{0pt}
    \item phonon emission;
    \item Friedel oscillations in an ordered sample;
    \item inhomogeneous broadening due to a disordered sample.
\end{enumerate}

\section*{Acknowledgments}
We thank Alexey Boyarsky for providing the data displayed in Fig. \ref{fig: beta decay spectra}. We are grateful to the referees for their helpful comments that helped improve this paper. We also wish to express our gratitude to Vladimir Gritsev for stimulating discussions.

\begin{appendix}
\numberwithin{equation}{section}

\section{Artificial Time Ordering}
\label{Artificial Time ordering}

Let us remark that $[\phi(0,t_{1}),\phi(0,t_{2})] = -i\pi\text{ sgn}(t_{1}-t_{2})$. Substituting this into the BCH formula gives us for $t_{2} > t_{1}$:

\begin{equation}
e^{\lambda_{1}\phi(0,t_{1})}e^{\lambda_{2}\phi(0,t_{2})} = e^{i\pi\lambda_{1}\lambda_{2}}e^{\lambda_{2}\phi(0,t_{2})}e^{\lambda_{1}\phi(0,t_{1})}.
\end{equation}

\noindent Using the Heaviside step function, $\omega$, we find

\begin{equation}
\begin{aligned}
\Pi_{A}(t,t_{1},t_{2}) \hspace{-1.9cm}&\hspace{1.9cm}:= \langle FS|e^{\frac{iV_{0}}{2\pi v_{F}}\phi(0,0) }e^{iA\phi(0,t_{1})}e^{-iA\phi(0,t_{2}+t)}e^{-\frac{iV_{0}}{2\pi v_{F}}\phi(0,t)}|FS\rangle\\
&= e^{-i\frac{AV_{0}}{2v_{F}}}\langle FS|e^{\frac{iV_{0}}{2\pi v_{F}}\phi(0,0) }e^{iA\phi(0,t_{1})}e^{-\frac{iV_{0}}{2\pi v_{F}}\phi(0,t)}e^{-iA\phi(0,t_{2}+t)}|FS\rangle\\
&= e^{-i\frac{AV_{0}}{2v_{F}}}\omega(t_{1}-t)\langle e^{\frac{iV_{0}}{2\pi v_{F}}\phi(0,0) }e^{iA\phi(0,t_{1})}e^{-iA\phi(0,t_{2}+t)}e^{-\frac{iV_{0}}{2\pi v_{F}}\phi(0,t)}\rangle\\
&\hspace{0.3cm}+ \omega(t-t_{1})\langle FS|e^{\frac{iV_{0}}{2\pi v_{F}}\phi(0,0) }e^{-\frac{iV_{0}}{2\pi v_{F}}\phi(0,t)}e^{iA\phi(0,t_{1})}e^{-iA\phi(0,t_{2}+t)}|FS\rangle\\
&= e^{-i\frac{AV_{0}}{2v_{F}}}\omega(t_{1}-t)\langle e^{\frac{iV_{0}}{2\pi v_{F}}\phi(0,0) }e^{iA\phi(0,t_{1})}e^{-iA\phi(0,t_{2}+t)}e^{-\frac{iV_{0}}{2\pi v_{F}}\phi(0,t)}\rangle\\
&\hspace{0.3cm}+ e^{i\pi(\frac{V_{0}}{2\pi v_{F}})^{2}}\omega(t)\langle FS|e^{-\frac{iV_{0}}{2\pi v_{F}}\phi(0,t)}e^{\frac{iV_{0}}{2\pi v_{F}}\phi(0,0)}e^{iA\phi(0,t_{1})}e^{-iA\phi(0,t_{2}+t)}|FS\rangle\\
&\hspace{0.3cm}+ \omega(t-t_{1})\omega(-t)\langle FS|e^{\frac{iV_{0}}{2\pi v_{F}}\phi(0,0)}e^{-\frac{iV_{0}}{2\pi v_{F}}\phi(0,t)}e^{iA\phi(0,t_{1})}e^{-iA\phi(0,t_{2}+t)}|FS\rangle\\
&= e^{-i\frac{AV_{0}}{2v_{F}}}\omega(t_{1}-t)\langle e^{\frac{iV_{0}}{2\pi v_{F}}\phi(0,0) }e^{iA\phi(0,t_{1})}e^{-iA\phi(0,t_{2}+t)}e^{-\frac{iV_{0}}{2\pi v_{F}}\phi(0,t)}\rangle\\
&\hspace{0.3cm}+ e^{i\pi(\frac{V_{0}}{2\pi v_{F}})^{2}}\omega(t)\omega(t_{1}-t_{2}-t)\langle e^{\frac{iV_{0}}{2\pi v_{F}}\phi(0,0) }e^{iA\phi(0,t_{1})}e^{-iA\phi(0,t_{2}+t)}e^{-\frac{iV_{0}}{2\pi v_{F}}\phi(0,t)}\rangle\\
&\hspace{0.3cm}+e^{i\pi(\frac{V_{0}}{2\pi v_{F}})^{2}}e^{i\pi A^{2}}\omega(t)\omega(t+t_{2}-t_{1})\langle FS|e^{-\frac{iV_{0}}{2\pi v_{F}}\phi(0,t)}e^{\frac{iV_{0}}{2\pi v_{F}}\phi(0,0)}\times\\
&\hspace{0.3cm}\times e^{-iA\phi(0,t_{2}+t)}e^{iA\phi(0,t_{1})}|FS\rangle\\
&\hspace{0.3cm}+ \omega(t-t_{1})\omega(-t)\omega(t_{1}-t_{2}-t)\langle e^{\frac{iV_{0}}{2\pi v_{F}}\phi(0,0) }e^{iA\phi(0,t_{1})}e^{-iA\phi(0,t_{2}+t)}e^{-\frac{iV_{0}}{2\pi v_{F}}\phi(0,t)}\rangle\\
&\hspace{0.3cm}+ e^{i\pi A^{2}}\omega(-t)\omega(t_{2}+t-t_{1})\langle FS|e^{\frac{iV_{0}}{2\pi v_{F}}\phi(0,0)}e^{-\frac{iV_{0}}{2\pi v_{F}}\phi(0,t)}e^{-iA\phi(0,t_{2}+t)}e^{iA\phi(0,t_{1})}|FS\rangle
\end{aligned}
\end{equation}

\begin{equation*}
\begin{aligned}
\hspace{-1.9cm}&= \Big{(}e^{-i\frac{AV_{0}}{2v_{F}}}\omega(t_{1}-t) + e^{i\pi(\frac{V_{0}}{2\pi v_{F}})^{2}}\omega(t)\omega(t_{1}-t_{2}-t) + \omega(t)\omega(t+t_{2}-t_{1})\omega(-t_{2}-t)\\
&\hspace{0.3cm}+ e^{i\frac{AV_{0}}{2v_{F}}}\omega(t)\omega(t_{2}+t) + \omega(t-t_{1})\omega(-t)\omega(t_{1}-t_{2}-t)+e^{i\pi A^{2}}\omega(t_{2}+t-t_{1})\omega(-t)\Big{)}\\
&\hspace{0.3cm}\times\langle FS|\mathcal{T}\big{(}e^{\frac{iV_{0}}{2\pi v_{F}}\phi(0,0) }e^{iA\phi(0,t_{1})}e^{-iA\phi(0,t_{2}+t)}e^{-\frac{iV_{0}}{2\pi v_{F}}\phi(0,t)}\big{)}|FS\rangle\\
&=: f_{A}(t,t_{1},t_{2})\langle FS|\mathcal{T}\big{(}e^{\frac{iV_{0}}{2\pi v_{F}}\phi(0,0) }e^{iA\phi(0,t_{1})}e^{-iA\phi(0,t_{2}+t)}e^{-\frac{iV_{0}}{2\pi v_{F}}\phi(0,t)}\big{)}|FS\rangle.
\end{aligned}
\end{equation*}

\end{appendix}

\bibliography{BiBTeX_File.bib}
\end{document}